\documentclass[pra,aps,twocolumn,showpacs]{revtex4}
\usepackage{amsmath,graphicx,epsfig}
\usepackage{graphics}
\usepackage{setspace}
\usepackage{multirow}
\usepackage{amssymb}
\usepackage{amsfonts}
\usepackage{amssymb}
\usepackage{amsmath}
\usepackage{latexsym}
\usepackage{wasysym}
\usepackage{array}
\usepackage{paralist}
\usepackage{enumerate}
\usepackage{multirow}
\usepackage{rotating}
\usepackage{booktabs}
\usepackage{longtable}
\usepackage{natbib}
\usepackage{color}

\def\fig_width{3.375 in} 
\def\fig_width{3. in} 
\draft

\newlength{\defbaselineskip}
\newcommand{\setlinespacing}[1]%
           {\setlength{\baselineskip}{#1 \defbaselineskip}}

\begin{document}

\title{{Optical control of resonant light transmission for an atom-cavity system}}

\author{Arijit Sharma}
\affiliation{Raman Research Institute, Light and Matter Physics, Sadashivanagar, Bangalore 560080, India}
\author{Tridib Ray}
\affiliation{Raman Research Institute, Light and Matter Physics, Sadashivanagar, Bangalore 560080, India}
\author{Rahul V. Sawant}
\affiliation{Raman Research Institute, Light and Matter Physics, Sadashivanagar, Bangalore 560080, India}
\author{G. Sheikholeslami}
\affiliation{Raman Research Institute, Light and Matter Physics, Sadashivanagar, Bangalore 560080, India}
\author{D. Budker}
\affiliation{Department of Physics, University of California, Berkeley, California 94720-7300}
\affiliation{Helmholtz Institut, Johannes Gutenberg Universit\"at, 55099 Mainz, Germany}
\author{S. A. Rangwala}
\email{sarangwala@rri.res.in}
\affiliation{Raman Research Institute, Light and Matter Physics, Sadashivanagar, Bangalore 560080, India}
\date{\today}


\begin{abstract}

We demonstrate the manipulation of transmitted light through an optical Fabry-Perot cavity, built around a spectroscopy cell containing enriched rubidium vapor. Light resonant with the $^{87}$Rb D$_{2}$ ($F=2/F=1$) $\leftrightarrow F'$ manifold, is controlled by transverse intersection of the cavity mode by another resonant light beam. The cavity transmission can be suppressed or enhanced depending on the coupling of atomic states due to the intersecting beams. The extreme manifestation of cavity mode control is the precipitious destruction (negative logic switching) or buildup (positive logic switching) of the transmitted light intensity, on intersection of the transverse control beam with the cavity mode.  Both the steady state and transient response are experimentally investigated. The mechanism behind the change in cavity transmission is discussed in brief. 
 
\end{abstract}
\maketitle

\section{Introduction}

Resonantly coupled atom-cavity systems~\cite{SieBk,SalehBk} exhibit absorption~\cite{WangOptLett}, dispersion~\cite{WuOptLett}, scattering~\cite{MotschNJP}, and non-linear effects~\cite{WangPRL,YangPRA}. These processes occur due to the interaction of the electro-magnetic field supported by the cavity mode, with the intra-cavity medium. An important class of atom-cavity experiments are performed with cavities built around spectroscopic cells, representative of the weak coupling limit. The cavity can either be a ring (travelling wave) cavity~\cite{WuPRA,WuPRA80,Gea-BanaclochePRA} or a standing-wave cavity~\cite{YuPRA}. \\

 The relatively modest experimental requirements in the setting up of an atom-cavity system around spectroscopic cells makes this an attractive experimental platform. Such systems have been used to investigate bistability~\cite{revLugiatoGibbs,HMGibbsBk,OrozcoPRA,JGrippPRA,LowenauPRL,ATredicucciPRA,XueAPL,MlynekPRA,JoshiPRA,WangPRA65}, switching~\cite{BrownAPL}, and non-linear processes~\cite{WangPRA65051802}. The participating atomic level system offers diversity in the phenomena observed with such systems.\\

In the present work, the control of the transmission of resonant light through a Fabry-Perot cavity, built around an $^{87}$Rb vapor cell, is studied. The cavity mode intensity is altered using control light from the side, which intersects the cavity mode. The control results in either suppression (negative logic) or enhancement (positive logic) of the cavity mode intensity, due to the specific atomic states addressed by the control light. The intersecting control beam requires no specific alignment apart from a robust intersection of the cavity mode. Both steady state and transient changes in the transmitted intensity are studied and the results are interpreted qualitatively. While a detailed exploration of the physics behind the observered processes is the subject of an accompanying manuscript~\cite{rahul}, the discussion section of this paper outlines the qualitative interpretation for the experiments. The potential for the present experimental system as an all-optical switch is discussed. 

\section{Experimental Arrangement}
\subsection{Cavity Assembly}
 A symmetric sub-confocal Fabry-Perot (FP) cavity is constructed around a cylindrical spectroscopy cell (Triad Technologies Inc.) that is 75 mm in length and 25 mm in diameter, containing an isotopically enriched sample of $^{87}$Rb. The cell does not contain any buffer gas and does not have anti-relaxation wall coating. The cavity is constructed with high-reflectivity concave mirrors (Linos) of radius of curvature 250 mm, separated by $\approx$80 mm. In order to construct the cavity, one concave mirror is directly glued to one of the end faces of the cell, while the other mirror is glued onto an annular piezoelectric transducer of $\approx$3 mm thickness, which is in turn glued onto the opposite face of the cell.  
\begin{figure}
\center
\includegraphics[width=8.5 cm]{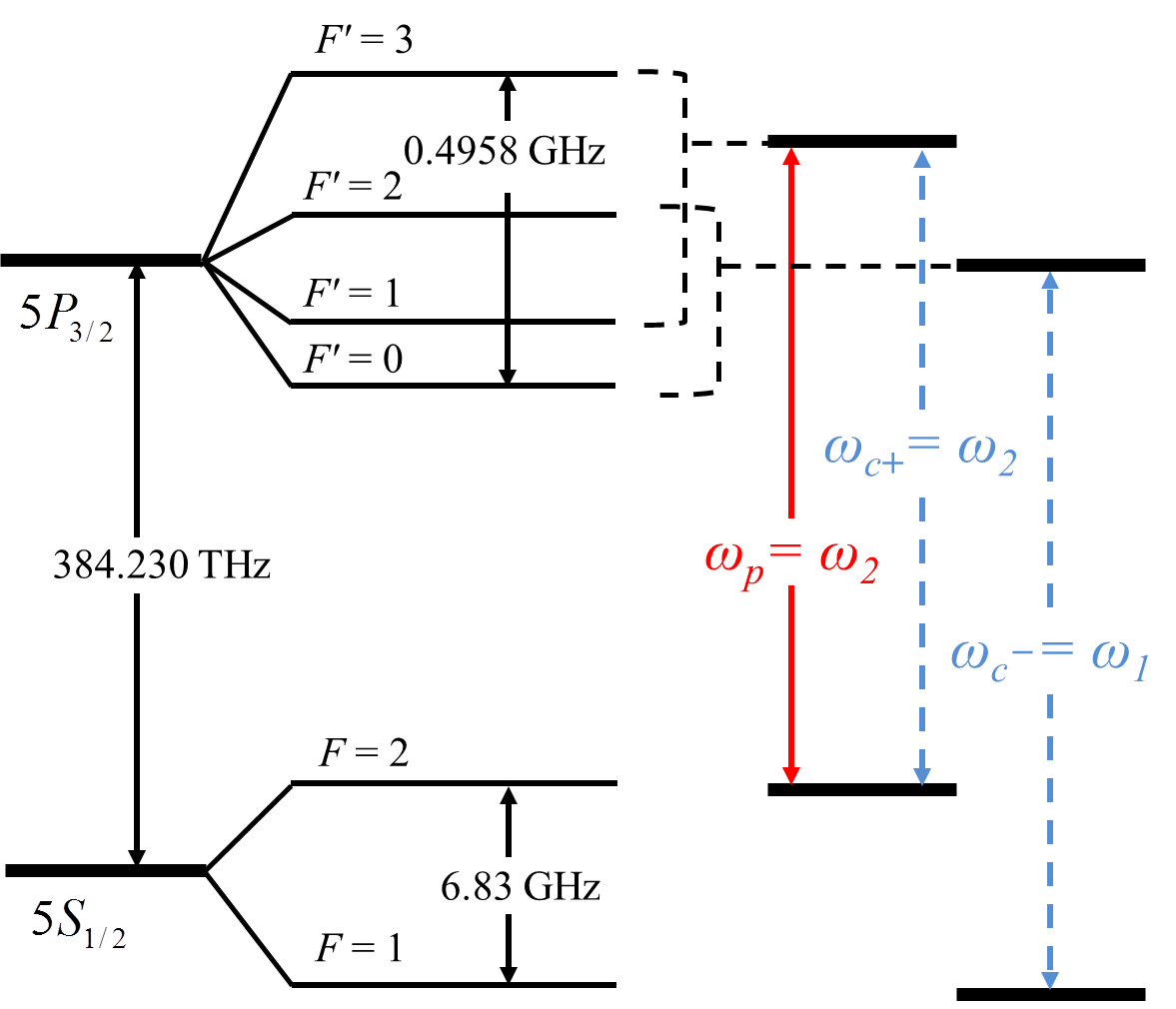}
\caption{(color online) The energy level schematic corresponding to the $D_2$ transitions of  $^{87}$Rb is shown. The right pane of the figure shows the reduced four-level scheme relevant to the experiment. The probe and the control laser frequencies $\omega_p$ and $\omega_c$ are shown. The superscripts '$+$' and '$-$' refers to the positive and negative logic of switching respectively.}
\label{Fig:RbSpectra3}
\end{figure}

\subsection{Optical System Setup}
\begin{figure}
\center
\includegraphics[width=8.5 cm]{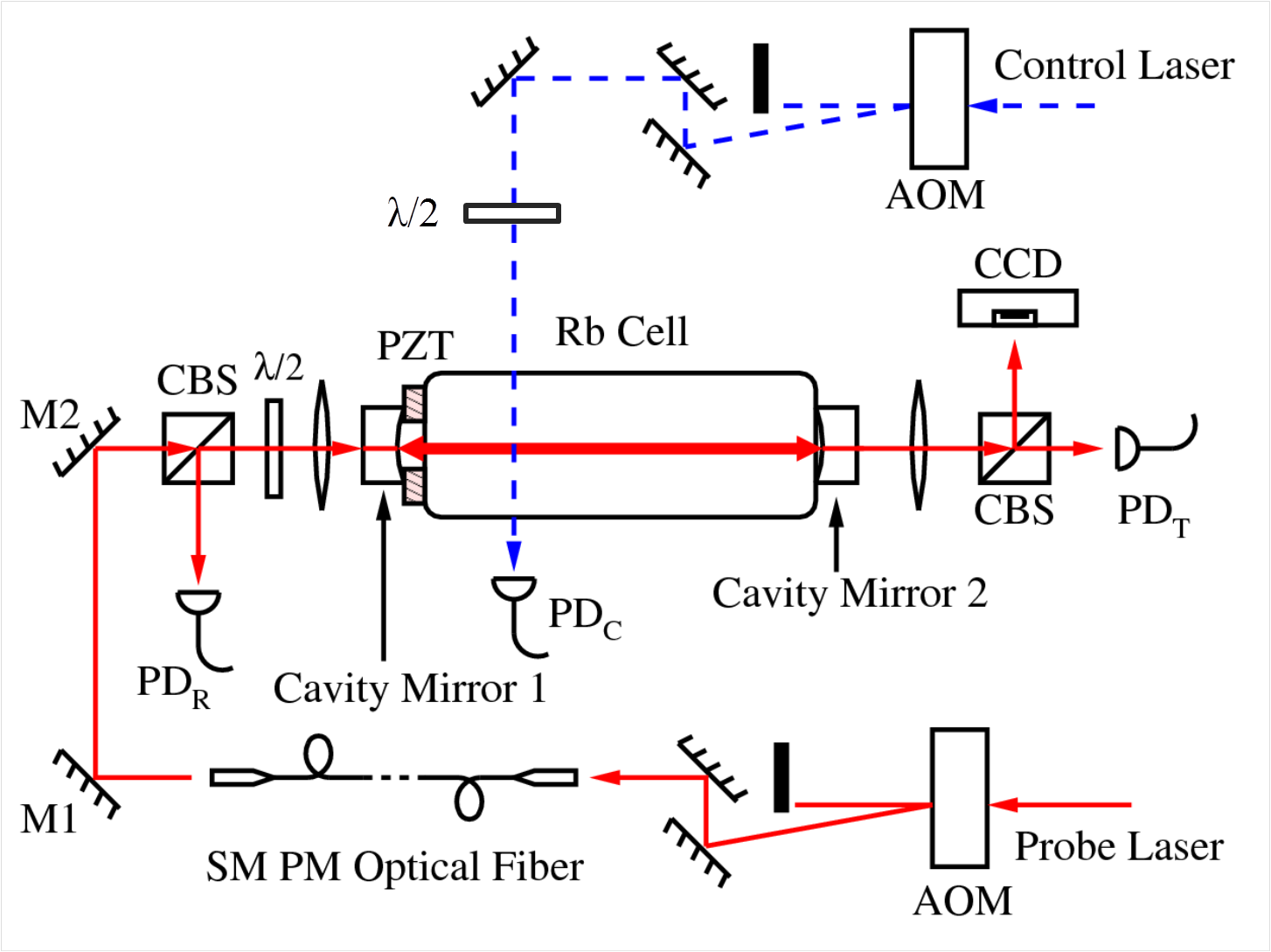} 
\caption{(color online) Schematic of the experimental setup. The FP cavity with the $^{87}$Rb cell is shown. The probe beam (solid line) is coupled into the cavity and the transmission and reflection ports are monitored using the photodiodes PD$_{T}$ and PD$_{R}$ respectively. The spatial profile of the transmitted mode is captured using a CCD camera. The control beam (dashed line) intersects the cavity mode over a small spatial region and is monitored using PD$_{C}$. Switching and intensity modulation of the two beams are done using AOM's as shown.}
\label{Fig:CavityWithRbCell}
\end{figure}

The energy level diagram relevant to the experiments reported is shown in Fig.~\ref{Fig:RbSpectra3}. At room temperature, the excited 5$^{2}$P$_{3/2}(F')$ manifold of $^{87}$Rb is unresolved due to Doppler broadening. The results are two absorption features, separated by $\approx$6.8 GHz.  $\omega_1$ and $\omega_2$ are identified as the frequencies corresponding to the Doppler absorption maxima of $F=1 \leftrightarrow F'$ and  $F=2 \leftrightarrow F'$ manifolds respectively. The laser beam coupled into the cavity mode is the {\it probe} laser beam, with frequency $\omega_{p}\leftrightarrow\omega_2$, and intensity $I_{p}$. The laser beam intersecting the probe laser beam in the cavity mode is the {\it control} laser beam, with frequency $\omega_{c}$ ($\omega_1$ or $\omega_2$) and intensity $I_{c}$. The pump and probe beams are both derived from independent lasers, both lasers are temperature stabilized and have a linewidth of $\approx2$MHz. Neither laser is actively frequency stabilized. The cavity mode waist is $\approx$ 150 $\mu$m. The diameter of the control laser beam is $\leq$1 mm and so its intersection with  mode is localized.
 Intensities of both laser beams can be controlled using acousto-optic modulators (AOM). The linearly polarized probe laser beam is passed through a single-mode polarization maintaining fiber, to obtain a TEM$_{00}$ spatial profile. A mode-matching lens is used for efficiently coupling the probe light into the cavity. The polarization direction of the probe beam can be changed before coupling it into the cavity using a half-wave ($\lambda$/2) plate. The control laser beam intersects the probe beam in the cavity mode with a small volume of intersection. A schematic of the experimental setup is shown in Fig.~\ref{Fig:CavityWithRbCell}. A CCD camera (not shown in the schematic) is used to image the fluorescence from the cavity mode and the intersecting control beam. 

\section{Cavity With Probe}

The coupling of the probe light into the cavity is optimized while it is on atomic resonance $\omega_2$. With this arrangement, the measured finesse F, of the cavity for light on atomic resonance is F $\approx 100$, for the experiments described in this article. 

\begin{figure}[t,h]
\center
\includegraphics[width=8.5 cm]{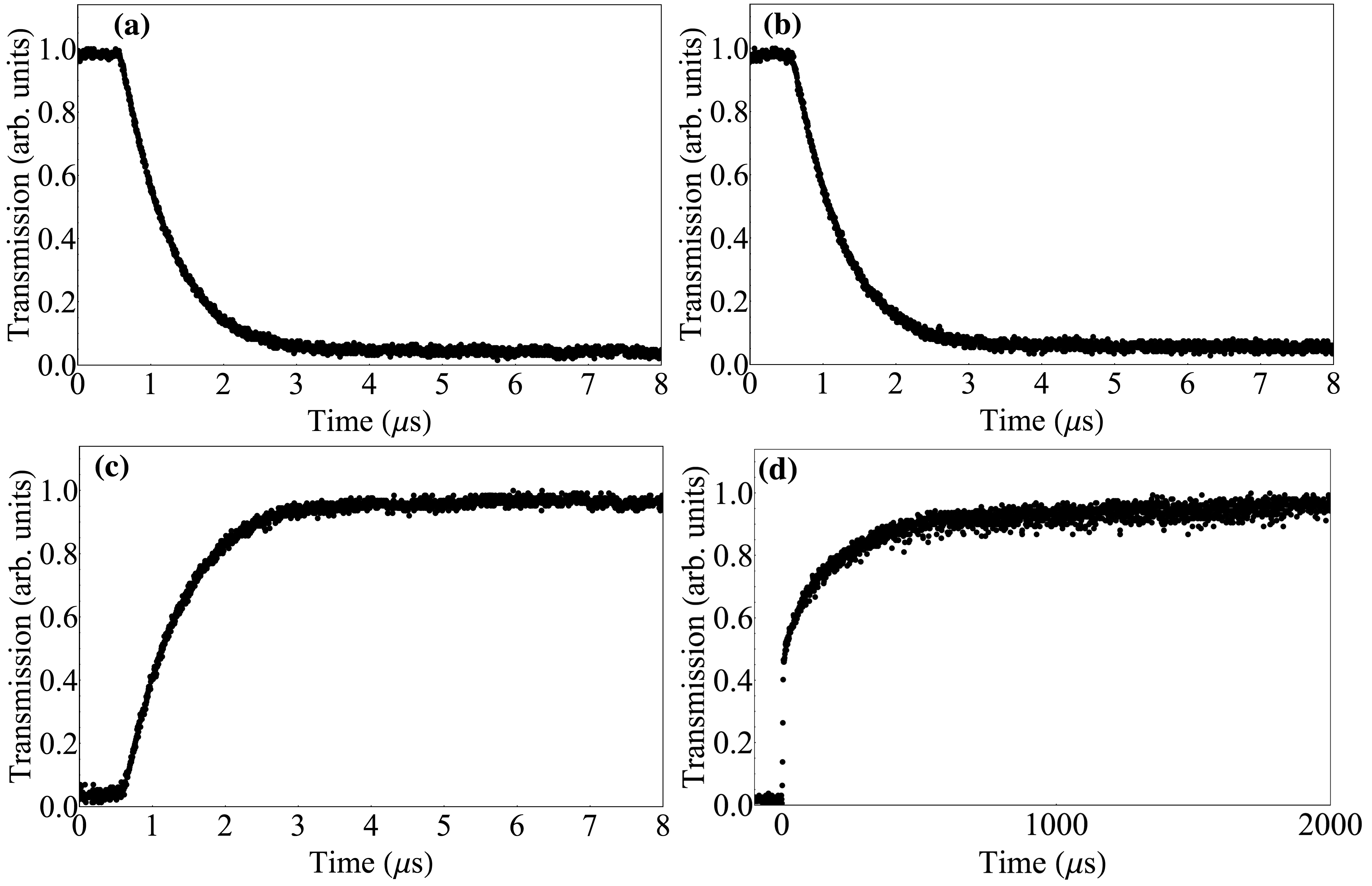}
\caption{(color online) The measured decay and rise of the cavity transmission for the probe, on- and off-resonance. The decay of the steady-state intensity, measured in transmission for the off-resonance case is (a), and the on-resonance case (b). In both cases the fall time has sub microsecond time constants. The rise in the transmitted intensity off- and on-resonance are shown in (c) and (d) respectively. Here the rise time constant for the off-resonant case is sub microsecond, whereas for the on-resonance case is significantly larger.}
\label{Fig:ProbeTimeConstants}
\end{figure}

Steady-state intensity in the cavity mode, while the probe laser is optimally coupled, can be maintained by adjusting the cavity length using the piezo electric transducer (PZT). The frequency of the probe laser is set to $\omega_2$ with an accuracy of a few MHz. The overall stability of the experiment is sufficient to allow many switching cycles to execute, without significant drift in the laser or cavity parameters. The transient behavior of probe light in the cavity mode is studied by rapid switching ON or OFF of the input probe beam, which is the first-order diffracted beam in the probe beam AOM. The rate of growth and decay of the probe light in the transmission port (PD$_T$) is shown in Fig.~\ref{Fig:ProbeTimeConstants}. The measurements are performed while the probe beam is tuned to atomic resonances $\omega_1$ and $\omega_2$, and at two OFF resonant frequencies $\omega'_1 \approx \omega_1 +2.5$ GHz and $\omega'_2 \approx \omega_2 - 1.3$ GHz. 

Figure ~\ref{Fig:ProbeTimeConstants} shows the on and off resonance buildup and decay of the probe light in the cavity. The time constants derived from single exponential fits to the data are presented in Table~\ref{Time_const_table_without_control_bare_cavity}. The measurements show that the decay of the light in the cavity mode is rapid ($\approx 1~\mu$s, both on and off resonance. In all cases the decay rates measured are limited by the response time of the photo-diode used. For the rise times, the off-resonant buildup of power in the cavity mode is rapid (measurement limited by the photo-diode response time), where as the on-resonance buildup time constant is $\approx60~\mu$s for $\omega_1$ and $\approx$ 180 $\mu$s for $\omega_2$. The large on-resonance rise time of the transmitted light and hence the intra-cavity intensity buildup rate can be directly related to the photon loss from the cavity mode due to resonant scattering of photons by the atoms. We observe that the time constants for establishment of the steady-state circulating intensity, and therefore steady transmission, are in the ratio of the transition strengths of the respective $^{87}$Rb atomic resonances. 

\begin{table}[b]
\caption{On- and off-resonance rise and fall time constants of the probe beam transmitted through the atom-cavity system. The reported time constants with standard deviation errors are derived from single exponential fits to the  transmitted PD$_T$ signals for the various cases. The control beam is off and the frequencies for the measurements are identified in the text and in Fig.~\ref{Fig:RbSpectra3}.}
\centering 
\begin{tabular}{ c c c c c } \\ [0.5ex]
\hline\hline 
Probe           & & Rise Time & & Fall Time \\ 
Frequency     & &($\mu{s}$) & &($\mu{s}$) \\ [0.25ex]
\hline 
$\omega'_1$ & & $0.75 \pm 0.03$ & & $0.66 \pm 0.06 $ \\ [0.25ex]
$\omega'_2$ & & $0.91 \pm 0.03$ & & $0.66 \pm 0.02 $ \\ [0.25ex]
$\omega_1$ & & $62 \pm 4$ & & $0.65 \pm 0.01 $ \\ [0.25ex]
$\omega_2$ & & $180 \pm 87 $ & & $0.65 \pm 0.02 $ \\ [0.25ex]
\hline \\
\end{tabular}
\textcolor{green}{\label{Time_const_table_without_control_bare_cavity}}
\end{table} 
\begin{figure}
\center
\includegraphics[width=7.5 cm]{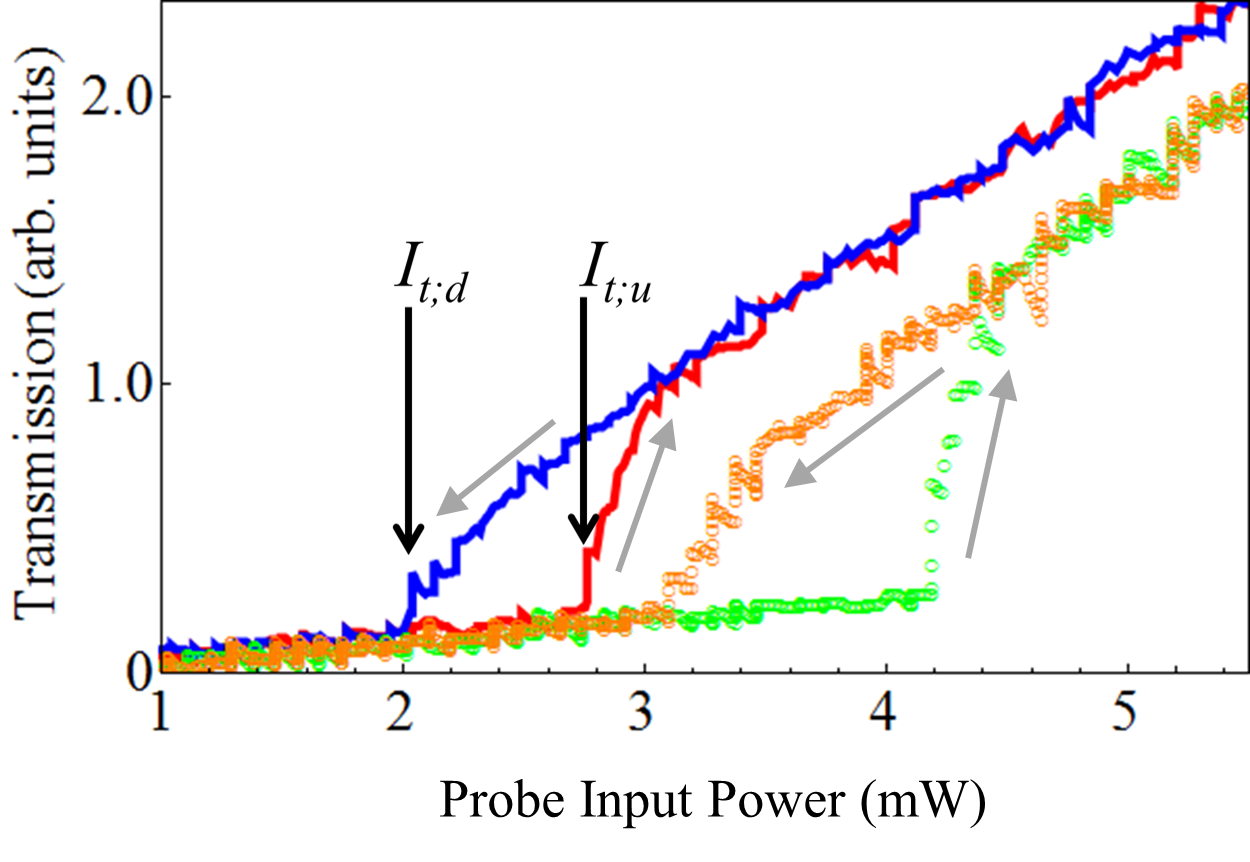}
\caption{(color online) The transmitted light through the cavity, at $\omega_2$, as a function of input light power for up and down cycle. The arrows shows the directions of the input intensity scan. The cavity transmission with increasing input power is the solid red line while the solid blue line gives the transmission as a function of falling input power, in the absence of the control beam. When the control beam is tuned to $\omega_1$, the hysteresis feature shifts to higher input powers. In the presence of the control light with a power of 10 mW, the new increasing power response is represented by the green circles, while the orange circles give the falling input-power response. The shift in the cavity response, in the presence of the control light results in the control of the intra-cavity light intensity.}
\label{Fig:Hysteresis}
\end{figure}

On atomic resonance, the probe light transmitted through the cavity exhibits bistable behaviour with respect to the input light intensity. This is observed over the entire inhomogenously broadened resonance, when the cavity length is adjusted to support transmission. The input probe intensity into the cavity is ramped from low to high (up cycle) and back from high to low (down cycle), by using the AOM in the path. The evolution of the cavity transmitted power (PD$_T$) as a function of the incident probe power (PD$_I$) exhibits bistable behavior, as shown in Fig.~\ref{Fig:Hysteresis}. The lack of active stabilization of the cavity length as well as the laser frequency leads to some scatter in the cavity input-output traces. To overcome this, a seven-point moving average is implemented on the data to obtain the traces presented in Fig.~\ref{Fig:Hysteresis}. This results in the blunting of the sharp threshold features, but nevertheless illustrates the sharp onset of transmission. 

In the up cycle, the transmission intensity shows a relatively small linear increase for input light intensity $I<I_{t;u}$. At $I_{t;u}$ a small change in input intensity results in a sharp increase of intra-cavity light buildup and cavity consequently transmits, exhibiting the characteristics of threshold behaviour. Beyond this point, the transmitted light intensity grows linearly with the input intensity, with significantly larger slope than the response below the threshold. In the down cycle, as the input intensity is gradually reduced, the above-threshold linear regime persists well below $I_{t;u}$, until $I_{t;d}$ below which the transmitted intensity follows the behaviour of $I<I_{t;u}$. The intensity buildup in the cavity exhibits hysteresis in the range $I_{t;d}<I<I_{t;u}$. 

\section{Cavity With Probe and Control}

With the cavity mode light frequency $\omega_p$ on atomic resonance, the addition of another beam of light on atomic resonance, which intersects the cavity mode can alter the intra-cavity mode intensity. Specifically, for $\omega_p=\omega_2$, when $\omega_c=\omega_2$ ($\omega_{c^+}$ in Fig.~\ref{Fig:RbSpectra3}), the same atomic transition, the hysteresis behaviour shifts to lower input light power. On the other hand when $\omega_c=\omega_1$ ($\omega_{c^-}$ in Fig.~\ref{Fig:RbSpectra3}), the hysteresis behavior shifts to higher input probe light power, as shown in Fig.~\ref{Fig:Hysteresis}. In the experiments reported below, we show both the rapid near extinction and growth of light in the cavity mode due to intersection with the appropriately tuned control laser beam. 

\subsection{Cavity Mode Extinction}
\begin{figure}
\center
\includegraphics[width=7.5 cm]{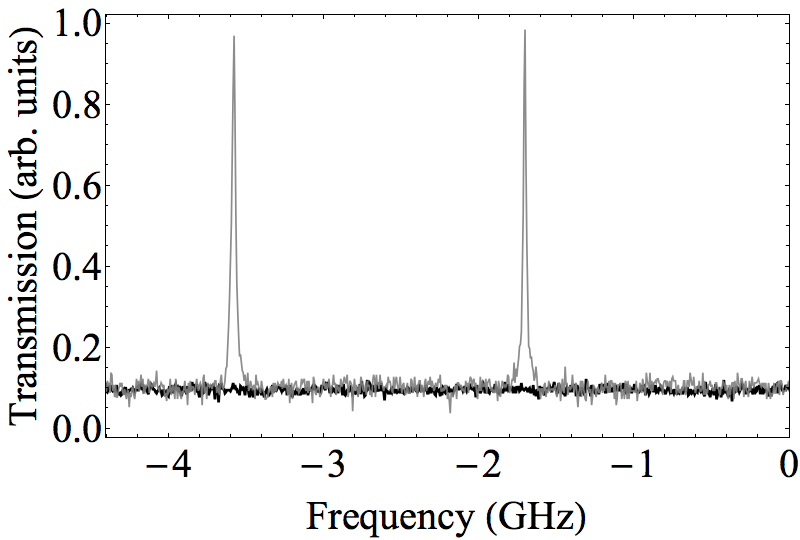}
\caption{The cavity transmission of probe light over a free spectral range, as a function of the cavity length (in frequency units), is illustrated by the thin  grey trace.  When the control light intersects the cavity mode, we see the complete suppression of the cavity transmission peaks shown as the thick black trace. The transmission has been extinguished and not been shifted in frequency.}
\label{Fig:CavityTransmission}
\end{figure}

To study the system for cavity-mode extinction, $\omega_p$ and $\omega_{c^-}$ are tuned to the $\omega_2$ and $\omega_1$, respectively. The choice of complementary transitions for the probe and control laser can be reversed, with no qualitative change in the experimental results. A weak beam of the control laser is sufficient to extinguish the transmission of $\omega_p$ completely as shown in Fig.~\ref{Fig:CavityTransmission}. 

In this case, the presence of the control beam shifts the entire hysteresis response to higher values of cavity input intensities, as seen in Fig.~\ref{Fig:Hysteresis}. Thus the presence of control light inhibits the transmission for any input light intensity that is below the new threshold. For input light intensity above the threshold, the control light causes partial attenuation.

On varying the polarizations of both beams it is found that changes in the relative polarizations of the probe and the control beam have no measurable effect on the attenuation of the cavity transmission. Only the intensity of the control beam determines the degree of attenuation, in the range of input intensities of $\omega_p$ over which the hysteresis response is seen. The light at the output of the cavity has the same polarization state as the input probe light. 

\begin{figure}
\center
\includegraphics[width=7.5 cm]{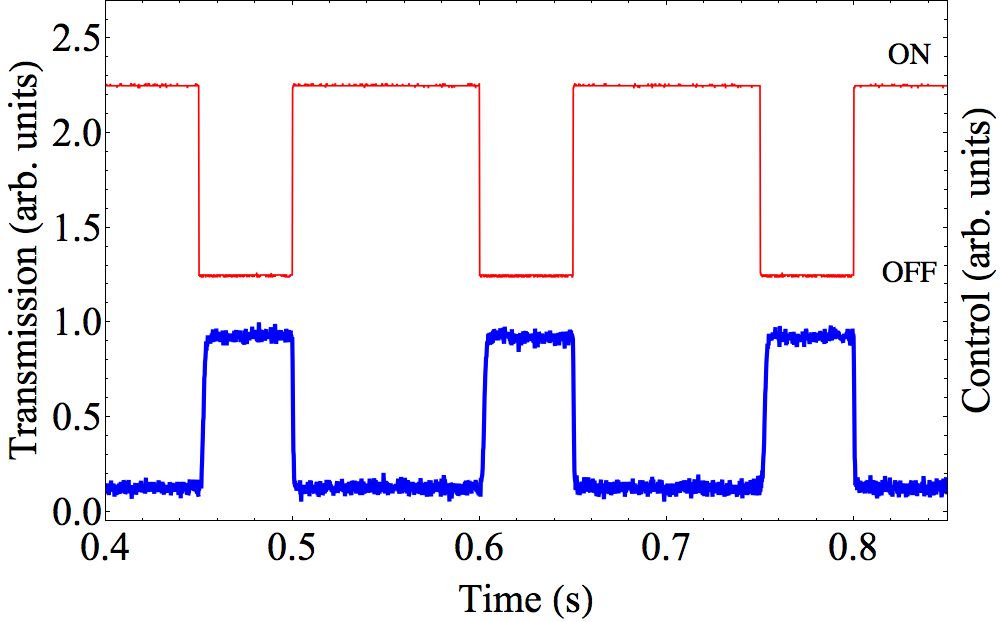}
\caption{A time trace recording three cycles of switching. The upper trace shows the temporal profile of the control laser and the bottom trace shows the temporal profile of the cavity transmission. The transmission is high when $I_c = 0$ and low when $I_c > I_{c;s}$. For this set of probe and control frequencies, the switch operates in negative logic.}
\label{Fig:Switching}
\end{figure}

The suppression of the cavity transmission of the probe beam by the control beam can be used to fabricate an all-optical negative-logic switch. The intensities of the probe and control beams are chosen so that cavity mode extinction is seen. The control beam is turned ON and OFF by means of the control AOM and the cavity output is monitored. When the control is ON, the probe transmission is dramatically reduced. The transmission goes back to the initial level when the control beam is turned OFF. The optical switching of the probe light, by the application of the control beam, over three cycles is shown in Fig.~\ref{Fig:Switching}, demonstrating high fidelity switching of cavity transmission.

\begin{figure}
\center
\includegraphics[width=7.5 cm]{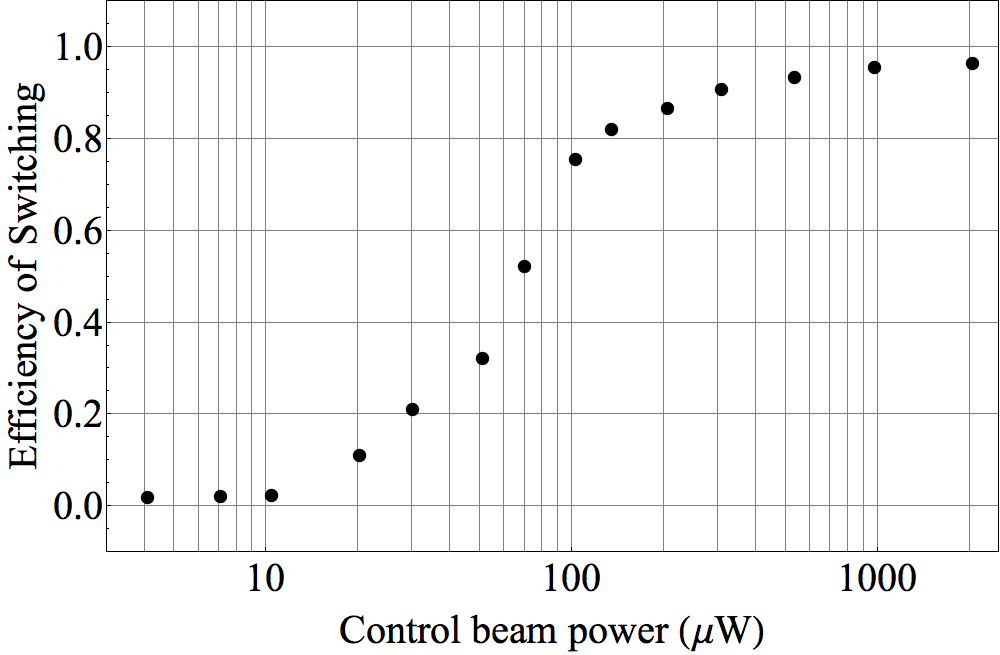}
\caption{The change in the efficiency of the switching, as a function control laser power. The $h_{off}$ value is baseline corrected for determining the efficiency.}
\label{Fig:SwitchingEfficiency}
\end{figure}

The efficiency with which the control beam can switch the probe transmission, is defined as $e = 1 - h_{on}/h_{off}$. Here $h_{off}$ and $h_{on}$ are the heights of the cavity transmission peaks, at the given control beam power, when it is turned OFF and ON, respectively. The transmission peak heights are measured with respect to the non resonant baseline light level, which remains the same irrespective of the state of the switching laser, as seen in Fig.~\ref{Fig:CavityTransmission}. In absolute terms for this measurement, the baseline light level is $\approx 10$ times lower than the peak heights, demonstrating the high fidelity of the digital all optical switch. The variation of the switching efficiency, as a function of the control laser intensity is measured using a fixed intensity $I_{p}$ of the probe laser where complete switching is achieved, and varying the intensity of the control laser $I_{c}$. As the control power increases beyond a threshold power, the efficiency of the attenuation increases rapidly until it reaches $e\approx0.8$, beyond which a further increase in efficiency is gradual. This is illustrated in Fig.~\ref{Fig:SwitchingEfficiency}.

\begin{figure}[b]
\center
\includegraphics[width=7.5 cm]{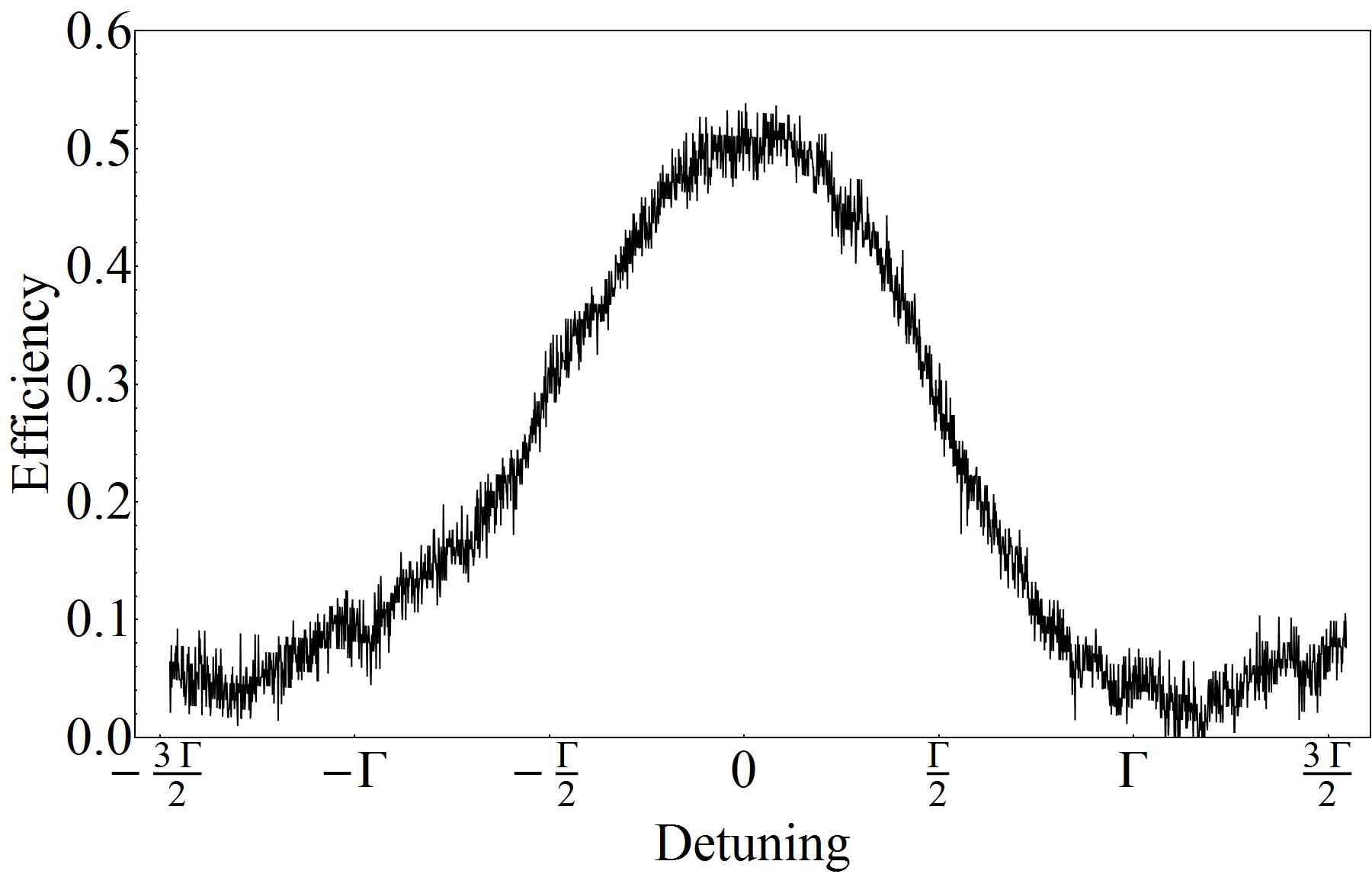}
\caption{The change in the efficiency of attenuation as a function of the control laser frequency across the Doppler absorption profile of the control laser.}
\label{Fig:Detuning}
\end{figure}

The control frequency dependence of the efficiency is investigated by setting the probe and control frequencies and intensities such that for $\omega_p = \omega_2$ and $\omega_{c^-} = \omega_1$, and $e\approx0.5$. The cavity length is adjusted to support the probe light buildup in the absence of the control beam. The control light is now frequency scanned across the full Doppler width of $\omega_1$ and the cavity output is monitored. From the result given in Fig.~\ref{Fig:Detuning}, we observe that the efficiency of attenuation decreases to zero $\approx \Gamma$ away from peak absorption, where $\Gamma\approx500$ MHz, is the full width at half maximum of the Doppler width of $\omega_1$ at room temperature. The observation confirms the role of resonant interaction in the attenuation of the cavity transmission.

\begin{figure}[h]
\center
\includegraphics[width=7.5 cm]{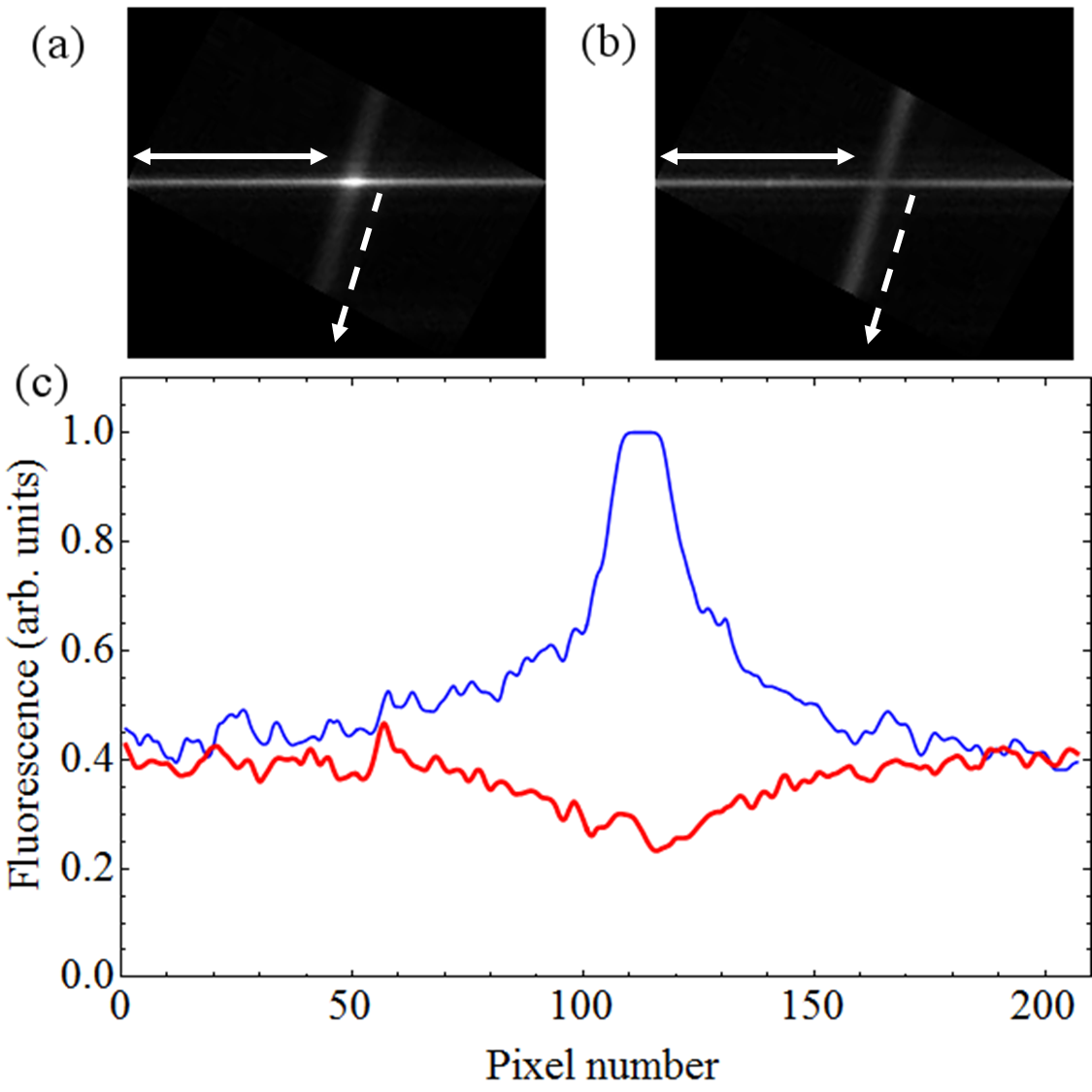}
\caption{The image of the intersection region of the cavity mode with the control beam, (a) and (b) (200 pixels $\approx$ 25 mm). The bold two-headed arrow marks the direction of the cavity mode and the dashed single headed arrow the control beam. The system is in the partial attenuation regime for this experiment. (a) illustrates increased light scattering accompanying probe attenuation, i.e. the two beams on complementary transitions and (b) decreased light scattering accompanying probe enhancement, i.e. the two beams on the same transition. The fluorescence intensity in arbitrary units along the cavity mode is illustrated in (c), from which it is clear that the scattering from the intersection volume is enhanced for the negative logic configuration (blue, thin line) and reduced for the positive logic case (red, thick line).}
\label{Fig:fluorescence}
\end{figure}

The atomic fluorescence from the cavity mode and the intersecting control beam is monitored during the switching. In the negative-logic partial switching case, we observe enhanced fluorescence in the intersection volume, as shown in Fig.~\ref{Fig:fluorescence} (a) and (c), when the cavity transmission is attenuated. The increased fluorescence is due to the pulling away of the atom-light system from saturation by $\omega_1$ in the region of overlap, resulting in increased absorption of cavity mode light locally.

\begin{figure}
\center
\includegraphics[width=7.5 cm]{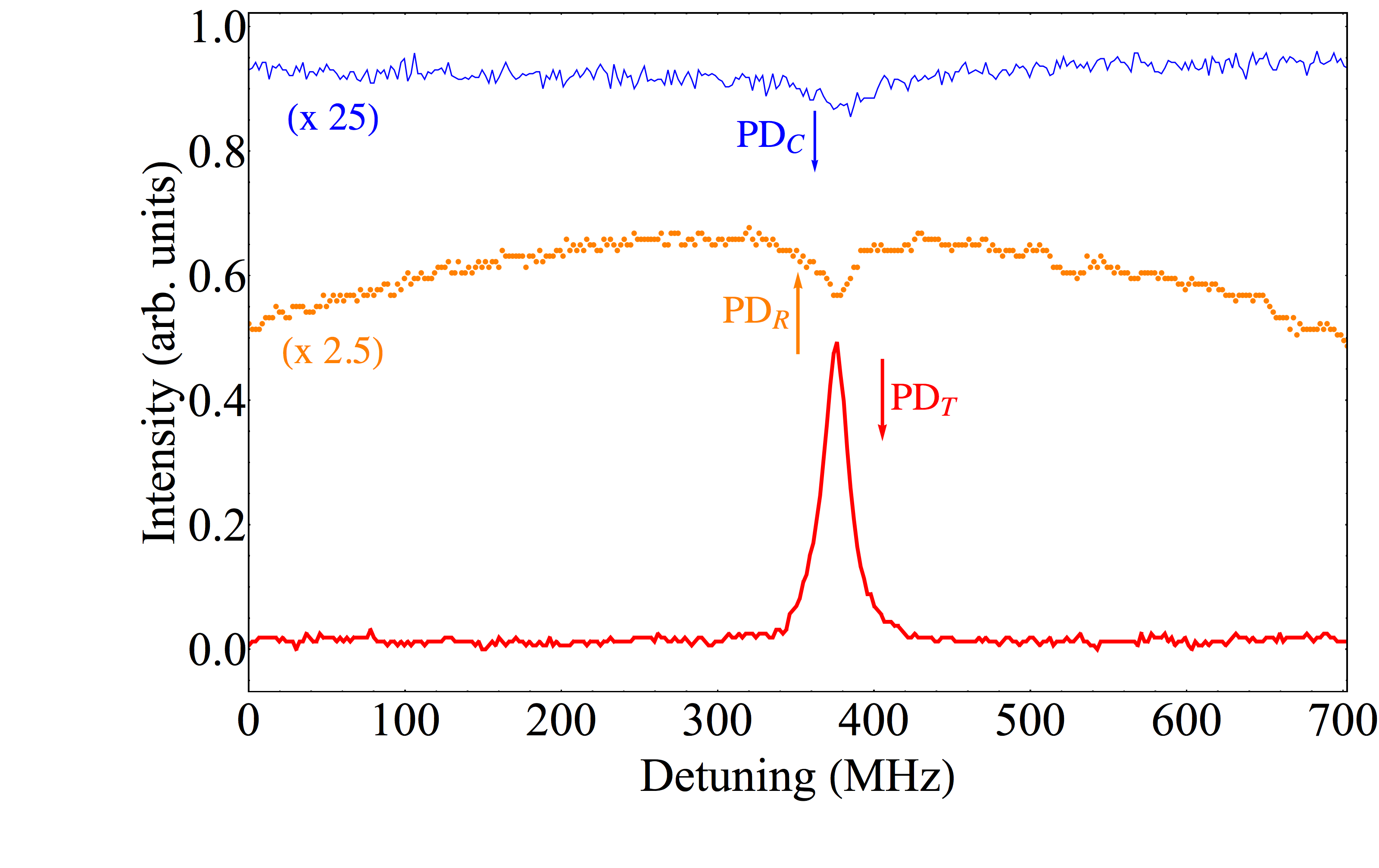}
\caption{A sweep across the cavity resonance is shown in the presence of the control beam. The orange (dotted) and red (thick line) traces are cavity reflection (PD$_{R}$) and transmission (PD$_{T}$) signals, respectively. The blue trace (solid line) is the transmitted intensity (PD$_{C}$) of the control beam. In the absence of the control light, the PD$_{R}$ dip is deeper and the PD$_{T}$ peak is higher. When the control light is turned ON, at the cavity resonance, the peak in PD$_{T}$ goes down ($\downarrow$), the dip in PD$_{R}$ decreases ($\uparrow$) and the transmitted control light goes down ($\downarrow$). The multiplicative factores indicate the relative signal strengths.}
\label{Fig:PartialAttenuationScan}
\end{figure}

The energy budget of the light intensity in the attenuation regime is understood by looking at the power in PD$_R$, simultaneously with the power in PD$_T$ and PD$_C$. In the absence of the control light during a cavity scan, a fraction of the incident probe light is transmitted as detected by PD$_T$, and the rest of the light is rejected, manifesting as dip in the PD$_R$ signal, when the cavity transmits. In the presence of the control light, as we scan the cavity across the resonance, the transmission peak in PD$_T$ decreases, the reflection dip in PD$_R$ decreases (more light is reflected back), and a temporally coincident dip appears in the control light power, measured in transmission. The reflected, transmitted and control light beam variations are illustrated in Fig.~\ref{Fig:PartialAttenuationScan}. The dip in the control beam intensity occurs due to a fraction of this light getting utilized in the generation of enhanced fluorescence at the intersection volume of the probe and control beam. These changes show that, when the cavity light is attenuated by the control beam, the input light is rejected by the cavity. In the negative-logic complete switching case, all fluorescence along the cavity mode disappears and the input probe light cannot transmit through the cavity.

\begin{figure}
\center
\includegraphics[width=6.5 cm]{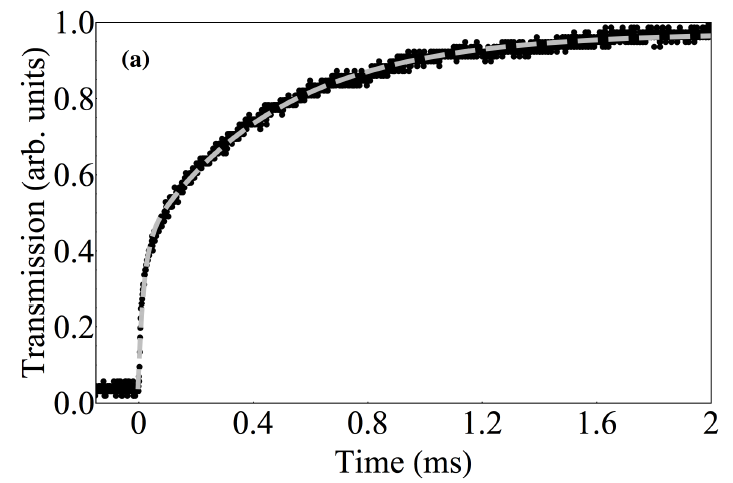}
\includegraphics[width=6.5 cm]{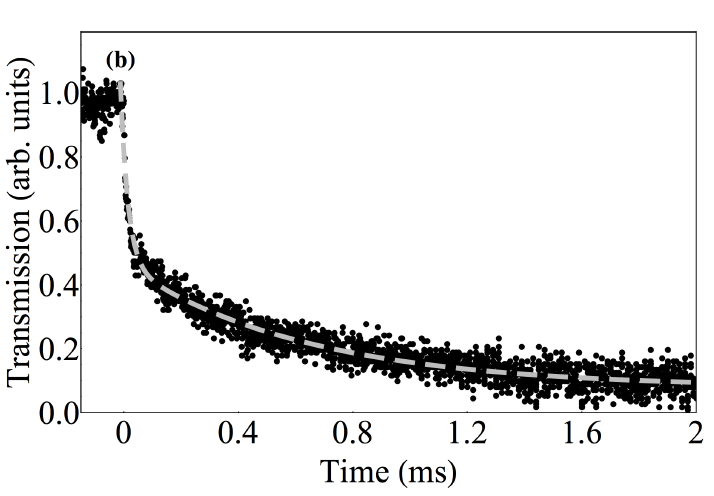}
\caption{The transient response of the cavity output probe light for the negative-logic switching. The rise and fall times are measured to be $\approx 1$ ms for 10 percent to 90 percent switching.}
\label{Fig:RiseFalltimeofSwitching}
\end{figure}

The transient behaviour of the switching action shown in Fig.~\ref{Fig:Switching}, is measured by tuning the probe intensity so that it is switched with high fidelity by the smallest control beam power. Once again $\omega_p$ and $\omega_c$ are tuned to $\omega_2$ and $\omega_1$, respectively. For the negative-logic switching, the transmission as the control beam is switched OFF and ON is shown in Fig.~\ref{Fig:RiseFalltimeofSwitching}. The transient behavior is not fit well by a single exponential. We find that a double exponential function fits the rise and fall behavior well and adding additional fit parameters  does not improve to the quality of the fit. The fast time constants are $\approx$ 25 $\mu$s and the slow time constants are $\approx$ 500 $\mu$s, as shown in Table~\ref{table:neglog}. The switching time of the control light AOM is less than 100 ns, and therefore does not limit our measurement. The reasons for the complex temporal response are addressed in the Discussion section.  

\begin{table}[h,b]
\caption{Rise and fall time constants with the standard deviations, for the negative-logic switching. The fast and slow time constants are obtained by double exponential fits to the data.}
\centering
\begin{tabular}{ c c c }
\hline\hline 
Case & Fast ($\mu{s}$) & Slow ($\mu{s}$)\\ [0.250ex]
\hline
Rise Time Constant & \textcolor{black}{46} $\pm$ \textcolor{black}{25} & \textcolor{black}{497} $\pm$ \textcolor{black}{139} \\ [0.5ex]
Fall Time Constant & \textcolor{black}{23} $\pm$ \textcolor{black}{10} & \textcolor{black}{570} $\pm$ \textcolor{black}{221} \\ [0.5ex]
\hline
\end{tabular}
\label{table:neglog}
\end{table} 

\subsection{Cavity Mode Enhancement}
Complementary to the extinction of light in the cavity mode described above, the light in the cavity mode can be enhanced by a transverse beam, demonstrating the switching of cavity mode in the positive logic. To do this the probe and control beam are adjusted to be on the same atomic resonance, $\omega_{p} = \omega_{c^+} = \omega_2$, while the experimental arrangement remains the same.  

For the enhancement case, the hysteresis response of the transmitted light intensity to the input probe light shifts to lower input light levels, in the presence of the control beam. This means that when the control beam is switched on, the cavity mode intensity in transmission increases.  

\begin{figure}
\center
\includegraphics[width=6.5 cm]{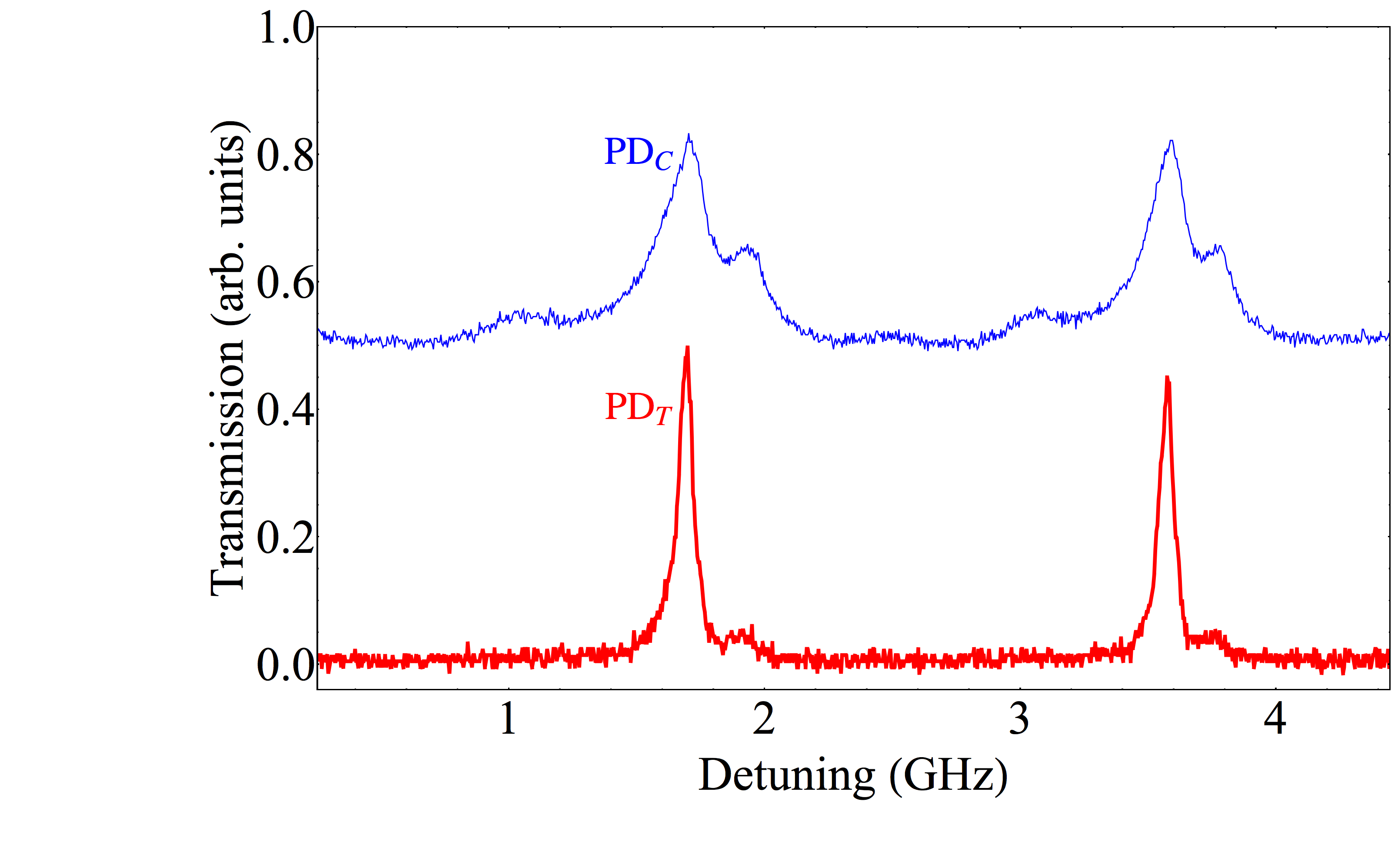}
\caption{Illustration of the cavity mode enhancement. The polarization direction of the control light is parallel to the cavity axis. In the panel, red (thick line) is the cavity transmission, blue (thin line) is the intensity of the control light ($\times30$ magnification), analyzed using a Wollaston prism. The increase in the intensity of the transmitted light for both the probe and control beams in the incident polarization state, when the cavity supports intensity buildup, results in a supression of fluorescence in the intersection volume, as seen in Fig.~\ref{Fig:fluorescence} (b) and (c). The cavity mode enhancement is independent of the control light polarization state.}
\label{Fig:PosLogicEnhan}
\end{figure}

A consequence of the buildup of cavity mode intensity, and therefore transmission, is seen when the input intensity of $\omega_p$ is adjusted to be marginally lower than $I_{t,u}$, so that the cavity mode intensity cannot be sustained and the transmitted light intensity is negligible. Turning on a weak control beam, intersecting the cavity mode, results in the buildup of the intensity in the cavity mode and the probe light transmits through the cavity. In steady state, the polarization of the transmitted cavity-mode light is identical to that of the input probe and is independent of the polarization state of the control beam. This is verified by rotating the plane of polarization of the control beam. Figure~\ref{Fig:PosLogicEnhan} illustrates the increase in the transmitted light intensity for both the probe and control beam, when the cavity is scanned across the resonance. In this case the fluorescence in the intersection volume of the probe and control beam decreases, as seen in Fig.~\ref{Fig:fluorescence} (b) and (c). Polarization analysis of the control light beam shows no change of polarization post transmission. Since the control light with this polarization cannot scatter along the cavity mode, we conclude that the intensity of the cavity transmitted light depends on the populations of the atomic states, which are intensity dependent, in the region of overlap of the probe and control beams. 

\begin{figure}
\center
\includegraphics[width=6.5 cm]{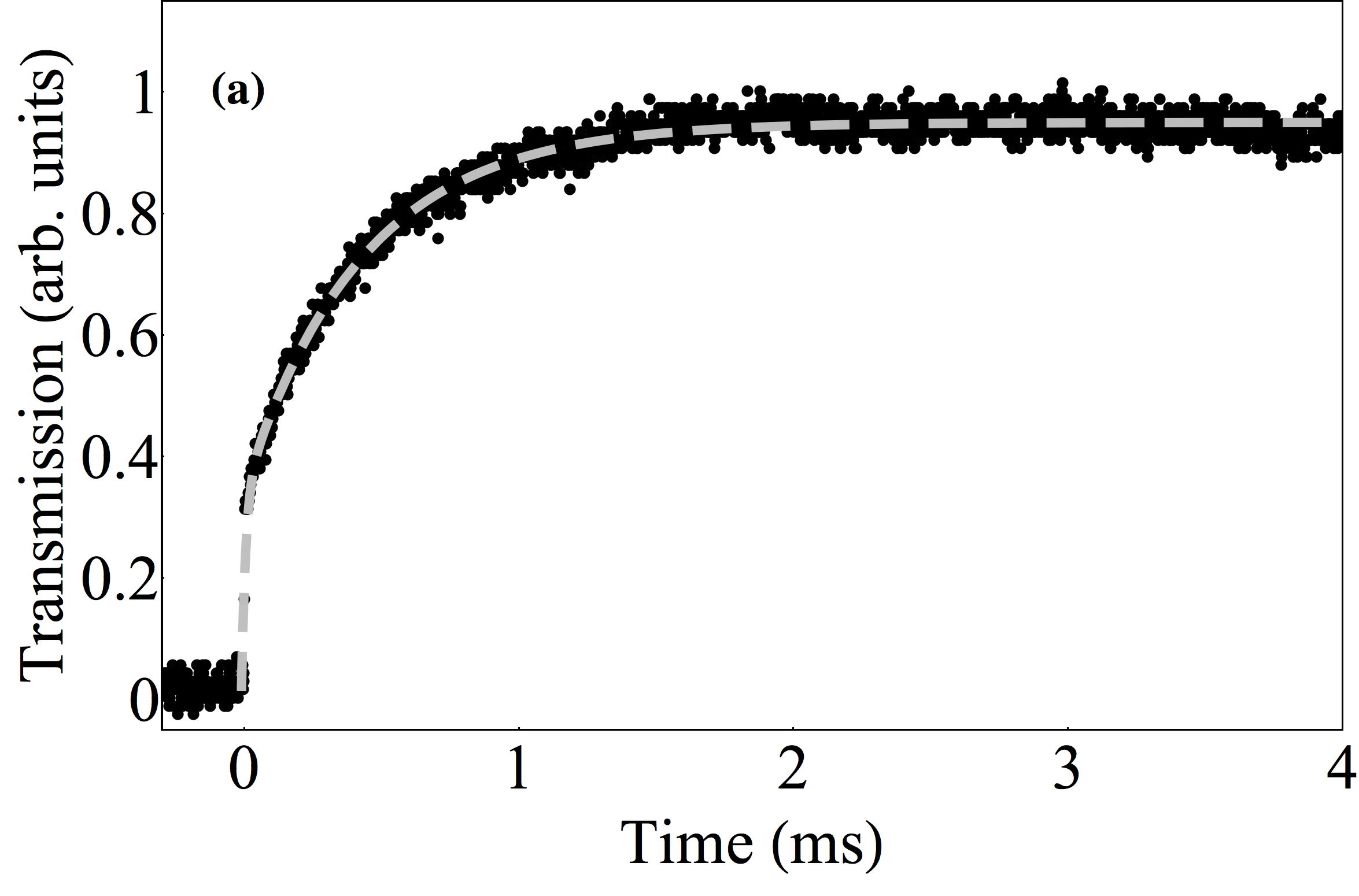}
\includegraphics[width=6.5 cm]{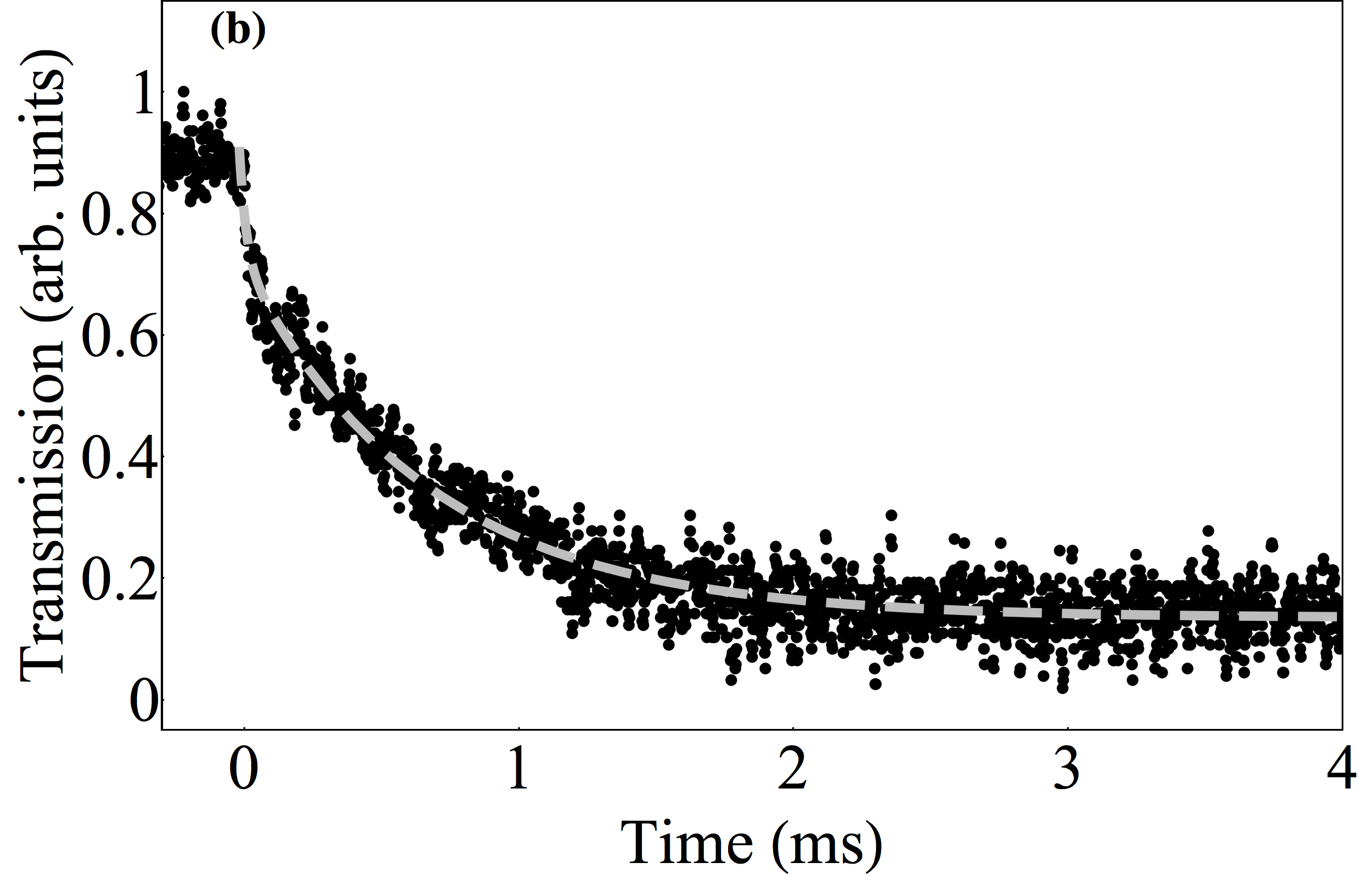}
\caption{ The transient response of the cavity output probe light for the positive-logic switching. Timescales of the fast and a slow responses obtained from double exponential fits to the data are provided in Table~\ref{table:poslog} and match closely to those of the negative logic case.}
\label{Fig:PosLogicTransient}
\end{figure}

\begin{table}[b]\label{table:poslog}
\caption{Rise and fall time constants with control beam for positive logic switching. The errors represent the mean value of the standard deviation, returned for the fast and slow time constants for least square double exponential fits, over several cycles of switching data.}
\centering
\begin{tabular}{ c c c }
\hline\hline
Case & Fast ($\mu{s}$) & Slow ($\mu{s}$)\\ [0.250ex]
\hline 
Rise Time Constant & \textcolor{black}{27} $\pm$ \textcolor{black}{11} & \textcolor{black}{505} $\pm$ \textcolor{black}{123} \\ [0.5ex]
Fall Time Constant & \textcolor{black}{24} $\pm$ \textcolor{black}{11} & \textcolor{black}{677} $\pm$ \textcolor{black}{128} \\ [0.5ex]
\hline
\end{tabular}
\end{table} 

A direct consequence of building up intensity in the cavity mode with a weak control beam intersecting the mode is that it allows the atom-cavity system to be utilized as a positive-logic all-optical switch. In this case, the cavity transmits the probe light when the control light is ON, and inhibits the transmission when the control beam is OFF. The experimental details for the measurement of the transient response for positive-logic switching remains the same as in the negative logic case. The responses at the rising and falling edges are illustrated in Fig.~\ref{Fig:PosLogicTransient} (a) and (b). As in the case of negative logic switching, the transmission response is best fitted by a double exponential. The fast and slow time constants provided in Table~\ref{table:poslog} match well with those of the negative logic case, given in Table~\ref{table:neglog}.

\section{Discussion}

The swiching and control of cavity beam transmission results from changes of the populations in the ground states for the atomic system. The experimental system here reduces to two  widely studied quantum systems. In the negative-logic case, the reduced transition system is a restricted, four-level system and for the positive-logic case, it is essentially a two-level system, with decay losses to cavity uncoupled states. Since the experiments are with a vapor cell, at room temperature and the size of the cell is much larger than the mode volume, the system can be treated as an open system in the grand canonical ensemble, where atoms continuously enter and exit the mode volume. A complete analytic and numerical description of such a system is complex, but a few careful simplifying assumptions can allow the generic system to be described with qualitative accuracy, as discussed in a separate manuscript~\cite{rahul}.

In the presence of the control beam, the atomic state populations of the Rb atoms in the intersection volume are altered relative to the rest of the cavity mode. When $\omega_p=\omega_{c^+}$, the photon loss from the cavity mode due to resonant scattering by the atoms reduces. The medium becomes more transparent to the probe light in the intersection volume, increasing the circulating intensity in the cavity mode. This in turn increases the saturation of the medium, resulting in increased  transmission of the input probe light, till a steady state energy density is established. The above discussion is consistent with the changed hysteresis conditions of the atom-cavity system.

When $\omega_p$ and $\omega_{c^-}$ are the set frequencies, the cavity transmission of the probe beam is suppressed. Here a fraction of the excited state  population is optically pumped back into resonance with $\omega_p$, resulting in the cavity mode atoms moving away from the saturation condition, creating more fluorescence and therefore radiative loss from the cavity mode, as seen in Fig.~\ref{Fig:fluorescence} (a). Due to the continuity of the probe light beam energy density across the whole cavity mode, the intensity in the entire mode decreases, till the steady state corresponding to the atom cavity response curves for this altered system is reached. 

The enhancement and suppression of the transmitted probe light in the atom-cavity system with the control beam can be used to implement a high fidelity all-optical switch, as shown in Fig.~\ref{Fig:Switching} and Fig.~\ref{Fig:RiseFalltimeofSwitching} for the case of negative logic, and in Fig.~\ref{Fig:PosLogicTransient} for positive logic. While this high fidelity is observed, the question of the response time for the change in state of the cavity mode needs explanation. The transients show a fast and slow time constant, for both negative and positive logic switching on the rising and falling edges. In both cases, the short time constants are measured to be a few 10 $\mu$s and the long time constants are in the range of $\approx 500~\mu$s. The atom flow affects the populations in the ground and excited states, both in the intersection volume and along the length of the cavity mode. Consequently, the state of transmission of the cavity changes in a complex manner, which is reflected in the temporal evolution of the cavity mode intensity~\cite{rahul}. Essentially the fast response comes from the light-atom interaction which exist in the cavity mode at the time of switching and the slow response is due to the flow of atoms in and out of the cavity mode.

An important future perspective of the study in this article is the interaction between laser cooled fluorescing atoms in a Magneto-Optical Trap (MOT) ~\cite{ELRaabPRL, CMonroePRL, CNCohen-TannoudjiPhysToday, WDPhillipsRevModPhys, MetcalfBk} and a cavity within which the MOT is created ~\cite{JYeAdvAtMolOptPhys, BermanBk, Wan12}. The optical control of atom-cavity interactions~\cite{Cha03,Her07,Wie10, Bie10,Wan12,Ray13} are of considerable importance for future experiments where fluorescing atoms in mixtures couple to cavity modes ~\cite{Ray14, Jyo14}. The present study therefore constitutes a preliminary, hot-vapor precursor to the corresponding cold vapor experiments. The motivation for performing these measurements on the D2 resonances for $^{87}$Rb stems from this connection to cold atomic vapour experiments, where $\omega_2$ maps to the cooling and $\omega_1$ maps to the repumping  transitions.  

\section{Conclusion}

In this work we demonstrate optical control of resonant light, in transmission through a standing wave cavity, built around hot atomic vapor. The power of control light required to affect the transmitted intensity of the resonant probe light through the cavity is shown to be a few hundred microwatt. The control light exhibits a threshold power, above which it alters the transmission properties of the cavity. The cavity-mode light can be extinguished by manipulating the control intensity, a fact which is used to demonstrate high-fidelity switching of transmitted light through this system. Both steady state and transient processes are experimentally characterized for this system. The physics of this system is discussed within the context of a driven multi-level quantum system~\cite{rahul}, and qualitative explanations for the observed steady-state and transient phenomena are provided. The robustness and ease of working with this system makes it suitable for exploitation as an all-optical switch, which operates with either positive or negative logic. The extension of this experimental technique to an ensemble of cold atoms contained within an cavity is a exciting possibility for probing atom-cavity coupling at ultracold temperatures. In this case, both collective strong coupling of the atoms to the cavity mode and the elimination of effects due to Doppler broadening will allow a number of possible experiments in the future. 

\section*{Acknowledgement}

We thank Seunghyun Lee, K. Ravi for help with the experiments, Byung Kyu Park for some preliminary analysis and Andal Narayanan for discussions.

\end{document}